\documentclass[aps,pra,onecolumn,groupedaddress,showpacs]{revtex4}

\usepackage{amssymb,amsmath}
\usepackage{graphicx}

\newcommand{\om}{\omega}

\newcommand{\pa}{\partial}

\begin{document}

\title{Two-component nonlinear wave of the cubic Boussinesq equation}

\author{G. T. Adamashvili}
\affiliation{Technical University of Georgia, Kostava str.77, Tbilisi, 0179, Georgia.\\ email: $guram_{-}adamashvili@ymail.com.$ }

\begin{abstract}In this work, we employ the generalized perturbation reduction method to find the two-component vector breather solution of the cubic Boussinesq equation $U_{tt}  - C U_{zz} - D U_{zzzz}+G (U^{3})_{zz}=0$. Explicit analytical expressions for the shape and parameters of the two-component nonlinear pulse
oscillating with the sum and difference of the frequencies and wave numbers are obtained.

\vskip+0.2cm
\emph{Keywords:} Two-component nonlinear waves, Generalized perturbation reduction method, Cubic Boussinesq equation.
\end{abstract}

\pacs{05.45.Yv, 02.30.Jr, 52.35.Mw}

\maketitle

\section{Introduction}

The nonlinear properties of shallow water waves can be modeled by various nonlinear partial differential equations. These include
the Benjamin-Bona-Mahony equation, the Korteweg-de Vries equation and different versions of the Boussinesq equations, among others [1-6]. There are various types of solutions that are revealed for these nonlinear equations. These solutions include the nonlinear solitary waves. Usually, two basic types of solitary waves are considered: single-component (scalar) and two-component (vector) nonlinear waves. Among single-component solitary waves, soliton and breather are considered quite often. In addition to single-component nonlinear waves, two-component waves, such as vector breathers, are also considered. Special type two-component vector breather oscillating with the sum and difference of the frequencies and wave numbers have been considered  using the generalized perturbative reduction method [1,7-11].

The generalized Boussinesq equation has the form [6,12]
\begin{equation}\label{eq1}
 \frac{\partial^{2} U}{\partial t^{2}}- C \frac{\partial^{2} U}{\partial z^{2}}  - D  \frac{\partial^{4} U}{\partial z^{4}}+ G   \frac{\partial^{2} U^{n}}{\partial z^{2}} =0.
\end{equation}
where $U(z, t)$ is a real function of space coordinate $z$ and time $t$ and represents the wave profile, while  $C,\;D $ and $G $ are arbitrary constants.

When $n = 3,$  Eq.(1) is reduced to the cubic Boussinesq equation
\begin{equation}\label{e2}
 \frac{\partial^{2} U}{\partial t^{2}}- C \frac{\partial^{2} U}{\partial z^{2}}  - D  \frac{\partial^{4} U}{\partial z^{4}}+ G   \frac{\partial^{2} U^{3}}{\partial z^{2}} =0,
\end{equation}
or in the dimensionless form
\begin{equation}\label{eq2}\nonumber
 \frac{\partial^{2} U}{\partial t^{2}}-  \frac{\partial^{2} U}{\partial z^{2}}  -   \frac{\partial^{4} U}{\partial z^{4}}+ 2  \frac{\partial^{2} U^{3}}{\partial z^{2}} =0.
\end{equation}

Eq.(2) is a well-known model of dispersive nonlinear waves which describes properties of nonlinear solitary waves in a one dimensional lattice and in shallow water under gravity.
The Boussinesq equation is also used in the analysis of many other phenomena [13, 14].
There are several well-known methods which can be used to solve the cubic Boussinesq equation. The tanh method, the variational iteration method, and several other
approaches to solve Eq.(2) and to analyze the solitary waves have been applied [12, 15, 16].

Note that, sometimes, another form of the cubic Boussinesq equation when the term ${\partial^{4} U}/{\partial z^{4}}$ is replaced by ${\partial^{4} U}/{{\partial z^{3}}{\partial t}}$  is also considered [17]
\begin{equation}\label{eq3}\nonumber
 \frac{\partial^{2} U}{\partial t^{2}}- C \frac{\partial^{2} U}{\partial z^{2}}  - D  \frac{\partial^{4} U}{{\partial z^{3}}{\partial t}}+ G   \frac{\partial^{2} U^{3}}{\partial z^{2}} =0.
\end{equation}

When the duration of the pulse $T>>1/\omega$ we can use the slowly varying envelope approximation [18-20]. In this
case we can represent the function $U(z, t)$ in the form
\begin{equation}\label{eq4}
U(z,t)=\sum_{l=\pm1}\hat{u}_{l}(z,t) Z_l,\;\;\;\;\;\;\;\;\;\;Z_l=e^{{il(k z -\om t)}}
\end{equation}
where $Z_{l}= e^{il(kz -\om t)}$ is the fast oscillating function, $\hat{u}_{l}$ are the slowly varying complex envelope functions, which satisfied inequalities
\begin{equation}\label{swa}
 \left|\frac{\partial \hat{u}_{l}}{\partial t}\right|\ll\omega
|\hat{u}_{l}|,\;\;\;\left|\frac{\partial \hat{u}_{l}}{\partial z
}\right|\ll k|\hat{u}_{l}|.
\end{equation}
$\omega$ and $k$ are the frequency and the wave number of the carrier wave. For the reality of $U$, we set: $ \hat{u}_{+1}= \hat{u}^{*}_{-1}$.

The purpose of the present work is to consider the two-component solution of the cubic Boussinesq equation (2) using generalized perturbative reduction method Eq.(8), when the function $U(z,t)$ satisfies the slowly varying envelope approximation, Eq.(3).

The rest of this paper is organized as follows: Section II is devoted to the linear part of the cubic Boussinesq equation for slowly varying complex envelope functions. In Section III, using the generalized perturbation reduction method, we will transform Eq.(2) to the coupled nonlinear  Schr\"odinger equations for auxiliary functions.
In Section IV, will be presented the explicit analytical expressions for the shape and parameters of the two-component nonlinear pulse. Finally, in Section V, we will discuss the obtained results.

\vskip+0.5cm
\section{The linear part of the cubic Boussinesq equation}

The linear part of the cubic Boussinesq equation (2) is given by
\begin{equation}\label{lin}
 \frac{\partial^{2} U}{\partial t^{2}}- C \frac{\partial^{2} U}{\partial z^{2}}  - D  \frac{\partial^{4} U}{\partial z^{4}} =0.
\end{equation}
We consider a pulse whose duration satisfies  the condition $T >>\omega^{-1}$.  Substituting  Eq.(3) into (5)  we obtain the connection between the parameters $\omega$ and $k$ in the form
\begin{equation}\label{dis}
 {\omega}^{2} = C  k^{2}- D k^{4}
\end{equation}
and the rest part of the linear equation
\begin{equation}\label{lin2}
\sum_{l=\pm1}Z_l [-2il\omega \frac{\partial \hat{u}_{l}}{\partial t}-2 i l k ( C  - 2  k^{2} D ) \frac{\partial \hat{u}_{l}}{\pa z}+\frac{\partial^{2} \hat{u}_{l}}{\partial t^{2}}
-( C - 6  k^{2} D )  \frac{\partial^{2} \hat{u}_{l}}{\partial z^2} -4 i l k D \frac{\partial^{3} \hat{u}_{l}}{\partial z^3} -D \frac{\partial^{4} \hat{u}_{l}}{\partial z^4}]=0.
\end{equation}

\vskip+0.5cm

\section{The generalized perturbative reduction method}

In order to consider the two-component vector breather solution of the Eq.(2), we use the generalized perturbative reduction method [1, 7-11] which makes it possible to transform the cubic Boussinesq equation for the functions $\hat{u}_{l}$ to the coupled nonlinear Schr\"odinger  equations for auxiliary functions $f_{l,n}^ {(\alpha)}$.
As a result, we obtain a two-component nonlinear pulse oscillating with the  difference and sum of the frequencies and wave numbers. In the frame of this method, the complex envelope function  $\hat{u}_{l}$ can be represented as
\begin{equation}\label{gprm}
\hat{u}_{l}(z,t)=\sum_{\alpha=1}^{\infty}\sum_{n=-\infty}^{+\infty}\varepsilon^\alpha
Y_{l,n} f_{l,n}^ {(\alpha)}(\zeta_{l,n},\tau),
\end{equation}
where $\varepsilon$ is a small parameter,
$$
Y_{l,n}=e^{in(Q_{l,n}z-\Omega_{l,n}
t)},\;\;\;\zeta_{l,n}=\varepsilon Q_{l,n}(z-v_{{g;}_{l,n}} t),
$$$$
\tau=\varepsilon^2 t,\;\;\;
v_{{g;}_{l,n}}=\frac{\partial \Omega_{l,n}}{\partial Q_{l,n}}.
$$

It is assumed that the quantities $\Omega_{l,n}$, $Q_{l,n}$ and $f_{l,n}^{(\alpha)}$ satisfies the inequalities for any $l$ and $n$:
\begin{equation}\label{rtyp}\nonumber\\
\omega\gg \Omega_{l,n},\;\;k\gg Q_{l,n},\;\;\;
\end{equation}
$$
\left|\frac{\partial
f_{l,n}^{(\alpha )}}{
\partial t}\right|\ll \Omega_{l,n} \left|f_{l,n}^{(\alpha)}\right|,\;\;\left|\frac{\partial
f_{l,n}^{(\alpha )}}{\partial \eta }\right|\ll Q_{l,n} \left|f_{l,n}^{(\alpha )}\right|.
$$

Substituting Eq.(8) into (7), for the linear part of the cubic Boussinesq equation (2)  we obtain
\begin{equation}\label{eqz}
\sum_{l=\pm1}\sum_{\alpha=1}^{\infty}\sum_{n=\pm 1}\varepsilon^\alpha Z_{l} Y_{l,n}[W_{l,n}
+\varepsilon J_{l,n} - \varepsilon^2 i l h_{l,n}  \frac{\partial }{\partial \tau}
-\varepsilon^{2} Q^{2} H_{l,n}\frac{\partial^{2} }{\partial \zeta^{2}}+O(\varepsilon^{3})]f_{l,n}^{(\alpha)}=0,
\end{equation}
where
\begin{equation}\label{cof}
W_{l,n}=- 2 n l\omega \Omega_{l,n}  + 2  n  l k Q_{l,n} ( C  - 2  k^{2} D ) - \Omega^{2}_{l,n}+( C - 6  k^{2} D )  Q_{l,n}^{2} - 4  n l k D Q_{l,n}^{3} - D Q_{l,n}^{4},
$$
$$
J_{l,n}=2i Q_{l,n} [l \omega  v_{{g;}_{l,n}}   -l k ( C  - 2  k^{2} D )   + n \Omega_{l,n}  v_{{g;}_{l,n}}   -( C - 6  k^{2} D )  n Q_{l,n} +6  l k D   Q_{l,n}^{2} + 2 D     n Q_{l,n}^{3}],
$$
$$
h_{l,n}=2(\omega + ln \Omega_{l,n}),
$$
$$
H_{l,n}=   C- v_{{g;}_{l,n}}^{2} - 6 D (   k +ln Q_{l,n} )^{2}.
\end{equation}

Equating to zero, the terms with the same powers of $\varepsilon$, from the Eq.(9) we obtain a series of equations. In the first order of $\varepsilon$, we have
a connection between of the parameters $\Omega_{l,n}$ and $Q_{l,n}$. When
\begin{equation}\label{fo}
 2 n l (  \omega   \Omega_{l,n}        -C  k  Q_{l,n}   + 2 D k^{3}   Q_{l,n}    + 2 D k   Q_{l,n}^{3}) +  \Omega_{l,n}^{2} -  C Q_{l,n}^{2} + 6 D  k^{2}   Q_{l,n}^{2}  + D Q_{l,n}^{4}=0,
\end{equation}
than
$$
f_{l,n}^{(1)}\neq0.
$$
From Eq. (10), in  the second order of $\varepsilon$, we obtain the equation
\begin{equation}\label{jo}
J_{l,n}=0
\end{equation}
and the expression
\begin{equation}\label{v}
v_{{g;}_{l,n}}=\frac{  l n    ( C k - 2 D  k^{2}  - 6 D  k  Q_{l,n}^{2} )  +  C Q_{l,n} - 6 D k^{2}  Q_{l,n}  -2 D  Q_{l,n}^{3}}{l n  \omega   +  \Omega_{l,n} }.
\end{equation}

In the third order of $\varepsilon$,  the linear part of the cubic Boussinesq equation (9), is given by
\begin{equation}\label{l}
\sum_{l=\pm1}\sum_{n=\pm 1}\varepsilon^{3} Z_{l} Y_{l,n}[ - i l h_{l,n}  \frac{\partial }{\partial \tau}- Q_{l,n}^{2} H_{l,n}\frac{\partial^{2} }{\partial \zeta^{2}}].
\end{equation}

Next we consider the nonlinear term proportional $Z_{+1}$ of the  cubic Boussinesq equation
\begin{equation}\label{non}
- 3 G[(k+Q_{+})^{2} ( | f_{+1,+1}^ {(1)}|^{2} +2 | f_{+1,-1}^ {(1)}|^{2} ) Y_{+1,+1} f_{+1,+1}^ {(1)}
$$$$
  +(k-Q_{-})^{2}   ( | f_{+1,-1}^ {(1)}|^{2}   +2   | f_{+1,+1}^ {(1)}|^{2} )Y_{+1,-1} f_{+1,-1}^ {(1)}].
\end{equation}

Combining Eqs.(14) and (15) we obtain the system of nonlinear equations
\begin{equation}\label{2eq}
  i \frac{\partial f_{+1,+1}^{(1)}}{\partial \tau} + Q_{+}^{2} \frac{H_{+1,+1} }{h_{+1,+1}} \frac{\partial^2 f_{+1,+1}^{(1)}}{\partial \zeta_{+1,+1} ^2}+\frac{3 G (k+Q_{+})^{2}}{ h_{+1,+1}}  ( | f_{+1,+1}^ {(1)}|^{2} + 2 | f_{+1,-1}^ {(1)}|^{2} ) f_{+1,+1}^ {(1)}=0,
$$$$
   i \frac{\partial f_{+1,-1 }^{(1)}}{\partial \tau} + Q_{-}^{2} \frac{H_{+1,-1} }{h_{+1,-1}} \frac{\partial^2 f_{+1,-1 }^{(1)}}{\partial \zeta_{+1,-1}^2}+\frac{3 G (k-Q_{-})^{2}}{h_{+1,-1}} ( |f_{+1,-1}^ {(1)}|^{2} +2 |f_{+1,+1} ^ {(1)}|^{2} )  f_{+1,-1}^ {(1)}=0.
 \end{equation}
where
\begin{equation}\label{qom}
 Q_{+}=Q_{+1,+1}= Q_{-1,-1},\;\;\;\;\;\;\;\;\;\;\;\;\;\;\;\;\;\;\;\;\;\;\;\;\;  Q_{-}=Q_{+1,-1}= Q_{-1,+1}.
 \end{equation}

\vskip+0.5cm
\section{The two-component vector breather of the cubic Boussinesq equation}

After transformation back to the variables $z$ and $t$, from Eqs.(16) we obtain the coupled nonlinear Schr\"odinger equations for the auxiliary functions $\Lambda_{\pm}=\varepsilon  f_{+1,\pm1}^{(1)}$ in the following form
\begin{equation}\label{pp2}
i (\frac{\partial \Lambda_{\pm}}{\partial t}+v_{\pm} \frac{\partial  \Lambda_{\pm}} {\partial z}) + p_{\pm} \frac{\partial^{2} \Lambda_{\pm} }{\partial z^{2}}
+q_{\pm}|\Lambda_{\pm}|^{2}\Lambda_{\pm} +r_{\pm} |\Lambda_{\mp}|^{2} \Lambda_{\pm}=0,
\end{equation}
where
\begin{equation}\label{OmQ}
p_{\pm}=\frac{ C- v_{\pm}^{2}  - 6 D (   k \pm  Q_{\pm} )^{2} }{2(\omega \pm \Omega_{\pm})},
$$
$$
 q_{\pm}=\frac{ 3 G (k\pm Q_{\pm})^{2}}{2(\omega \pm \Omega_{\pm})},
$$
$$
r_{\pm}=2 q_{\pm},
$$
$$
v_{\pm }= v_{g;_{+1,\pm 1}},
$$
$$
 \Omega_{+}=\Omega_{+1,+1}= \Omega_{-1,-1},\;\;\;\;\;\;\;\;\;\;\;\;\;\;\;\;\;\;\;\;\;\;\;\;\;  \Omega_{-}= \Omega_{+1,-1}= \Omega_{-1,+1}.
\end{equation}

The solution of Eq.(18) is given by [7-11]
\begin{equation}\label{ue1}
\Lambda_{\pm }=\frac{A_{\pm }}{b T}Sech(\frac{t-\frac{z}{V_{0}}}{T}) e^{i(k_{\pm } z - \omega_{\pm } t )},
\end{equation}
where $A_{\pm },\; k_{\pm }$ and $\omega_{\pm }$ are the real constants, $V_{0}$ is the velocity of the nonlinear wave. We assume that
$k_{\pm }<<Q_{\pm }$  and $\omega_{\pm }<<\Omega_{\pm }.$

Combining Eqs.(3), (8) and (20), we obtain the two-component vector breather solution of the cubic Boussinesq equation (2) in the following form:
\begin{equation}\label{vb1}
U(z,t)=\frac{2 }{b T}Sech(\frac{t-\frac{z}{V}}{T})\{  A_{+} \cos[(k+Q_{+}+k_{+})z
-(\omega +\Omega_{+}+\omega_{+}) t]
$$$$
 +A_{-}\cos[(k-Q_{-}+k_{-})z -(\omega -\Omega_{-}+\omega_{-})t]\}.
\end{equation}
The connections between parameters $A_{\pm },\; k_{\pm },\;\omega_{\pm },\;b$ and $T$  are given in Refs.[1, 7-11].

\vskip+0.5cm
\section{Conclusion}

In this paper, we study the two-component vector breather of the cubic Boussinesq equation (2) under the condition of the slowly varying envelope approximation.
 We consider nonlinear pulse with the width $T>>\Omega_{\pm }^{-1}>>\omega^{-1}$.
 Using the generalized perturbation reduction method Eq.(8), the Eq.(2) is  transformed to the coupled nonlinear Schr\"odinger equations (18) for the auxiliary functions $\Lambda_{\pm 1}$. As a result, the two-component nonlinear pulse oscillating with the sum and difference  of the frequencies and wave numbers Eq.(21), can be formed. The dispersion relation and the connection between parameters $\Omega_{\pm}$ and $Q_{\pm}$ are determined from Eqs.(6) and (11).

 Note that the two-component vector breather Eq.(21) we met in the different areas of physics and various nature of waves (see, for instance [1, 7-11, 21, 22] and references therein).

\end{document}